\documentclass[twocolumn,showpacs,prd]{revtex4}
\usepackage{mathrsfs,bm,multirow}
\usepackage{longtable,lscape}
\usepackage{txfonts}
\usepackage{amssymb}
\usepackage{indentfirst}
\usepackage{graphicx,booktabs}
\usepackage{color}
\usepackage{epstopdf}

\begin{document}

\title{Prediction of isoscalar charmoniumlike structures in the hidden-charm di-eta decays of higher charmonia}
\author{Dian-Yong Chen$^{1,3}$}
\email{chendy@impcas.ac.cn}
\author{Xiang Liu$^{1,2}$\footnote{Corresponding author}}\email{xiangliu@lzu.edu.cn}
\author{Takayuki Matsuki$^4$}
\email{matsuki@tokyo-kasei.ac.jp}
\affiliation{$^1$Research Center for Hadron and CSR Physics,
Lanzhou University $\&$ Institute of Modern Physics of CAS,
Lanzhou 730000, China\\
$^2$School of Physical Science and Technology, Lanzhou University,
Lanzhou 730000, China\\
$^3$Nuclear Theory Group, Institute of Modern Physics, Chinese
Academy of Sciences, Lanzhou 730000, China\\
$^4$Tokyo Kasei University, 1-18-1 Kaga, Itabashi, Tokyo 173-8602,
Japan}

\begin{abstract}

\if
Via the initial single chiral particle emission mechanism, we study the hidden-charm di-eta decays of charmoniumlike state $Y(4660)$ and the predicted charmonium $\psi(4790)$, i.e., $Y(4660)/\psi(4790) \to \eta\left[D^{(*)}\bar{D}^{(*)}\right]/\eta\ \left[D_s^{+(*)}{D}_s^{-(*)}\right]\to
J/\psi\eta\eta$ and answer to the question whether there exist isoscalar charmoniumlike structures. Our results indiate that there are enhancement structures near $D\bar{D}^*$, $D^*\bar{D}^*$ and $D_s\bar{D}_s^*$ thresholds in the $M_{\mathrm{max}}({J/\psi \eta})$ distributions of $Y(4660) \to \eta \eta J/\psi$. The calculation of $\psi(4790) \to \eta \eta J/\psi$ predicts the enhancement structures near $D^*\bar{D}^*$, $D_s\bar{D}_s^*$ and $D_s^*\bar{D}_s^*$ thresholds in the corresponding $M_{\mathrm{max}}({J/\psi \eta})$ distributions, which are accessible at future experiments, especially BESIII, Belle, BaBar and forthcoming BelleII.
\medskip
\fi

Considering the situation that a single chiral partilce, $\eta$, is initially emitted, we study the hidden-charm di-eta decays of charmoniumlike state $Y(4660)$ and the predicted charmonium $\psi(4790)$, i.e., $Y(4660)/\psi(4790) \to  J/\psi\eta\eta$ through the inetermediates, $\eta\left[D^{(*)}\bar{D}^{(*)}\right]$ and/or $\eta\ \left[D_s^{+(*)}{D}_s^{-(*)}\right]$, and answer to the important question whether there exist isoscalar charmoniumlike structures in the $D^{(*)}\bar{D}^{(*)}$ and/or $D_s^{+(*)}{D}_s^{-(*)}$ channels. Our results predict that there will be enhancement structures near $D\bar{D}^*$, $D^*\bar{D}^*$ and $D_s\bar{D}_s^*$ thresholds for $Y(4660)$ and near $D^*\bar{D}^*$, $D_s\bar{D}_s^*$ and $D_s^*\bar{D}_s^*$ thresholds for $\psi(4790)$ in the $M_{\mathrm{max}}({J/\psi \eta})$ distributions of $Y(4660)/\psi(4790) \to \eta \eta J/\psi$, respectively.
These peaks are accessible at future experiments, especially BESIII, Belle, BaBar and forthcoming BelleII.

\end{abstract}


\maketitle

The special behavior of the cross sections in the vicinity of thresholds has been noticed more than half a century ago in the nuclear reaction process \cite{Wigner:1948xx} based on unitarity in quantum mechanics and has been predicted to have the behavior $\sqrt{|s-s_{\rm th}|}$ near the threshold energy $s=s_{\rm th}$. Since then, the near threshold behavior has been studied by the various methods \cite{ Hategan:1976xx, Cabibbo:2004gq}. However, the extensive experimental studies also indicate a diversity of the threshold or cusp effects. In addition, the new experimental measurements of the charged $Z_b$ and $Z_c$ \cite{Belle:2011aa, Ablikim:2013mio} also show the {\it absence} of the enhancements near the thresholds of $B\bar{B}$ and $D\bar{D}$. These new phenomena stimulate us to propose a new approach to describe the behavior near the threshold. In Ref. \cite{Chen:2011pv}, we proposed a new mechanism, which is named as the initial-single-pion-emission (ISPE) mechanism to reproduce the lineshapes of the $\Upsilon(nS)\pi,\ \{n=1,3\}$ and $h_b(mP),\ \{m=1,2\}$ invariant mass distributions, where the $Z_b(10610)$ and $Z_b(10650)$ were discovered.

In 2011, the charged charmoniumlike structures near the
$D\bar{D}^*$ and $D^*\bar{D}^*$ thresholds were predicted in Ref.
\cite{Chen:2011xk} by studying the hidden-charm dipion decays of
higher charmonia and charmoniumlike states, where the
ISPE mechanism was adopted
\cite{Chen:2011pv}.This mechanism is described such that associating
an initially emitted one chiral particle, $\pi$ in this case, enhancement
in the invariant mass of $\psi'\pi$ can be seen through the triangle diagram with
charmonia and charmoniumlike states $\psi$ and $\psi'$ in the initial and final states.
Two years later, the BESIII Collaboration reported a
charged charmoniumlike structure $Z_c(3900)$ in $e^+e^- \to
\pi^+\pi^-J/\psi$ at $\sqrt{s}=4.26$ GeV \cite{Ablikim:2013mio},
which was confirmed by the Belle Collaboration \cite{Liu:2013dau}
and in Ref. \cite{Xiao:2013iha} later. The observation of
$Z_c(3900)$ confirms our prediction of a charged charmoniumlike
structure near the $D\bar{D}^*$ threshold existing in the
$J/\psi\pi^\pm$ invariant mass spectrum of $Y(4260)\to
J/\psi\pi^+\pi^-$  \cite{Chen:2011xk}, which provides a crucial test
of the ISPE mechanism.
{With enough experimental data, we have succeded in reproducing $Z_c(399)$
including background and final state interactions other than the ISPE diagrams \cite{Chen:2013coa}.
This paper clearly shows that the peak structures can be reconstructed mainly by the ISPE
mechanism even including all the effects, i.e., other diagrams and relative phases.
}

Besides these predictions listed in Ref. \cite{Chen:2011xk}, we have given
abundant phenomena of charged charmoniumlike structures by applying
the ISPE mechanism \cite{Chen:2013bha,Chen:2012yr} and the
initial-single-chiral-particle-emisssion (ISChE) mechanism
\cite{Chen:2013wca},  which is an extension of the ISPE mechanism.
The charged charmoniumlike structures with hidden-charm and
open-strange channels in the $J/\psi K^+$ invariant mass spectrum for the
processes $\psi(4415)/Y(4660)/\psi(4790)\to J/\psi K^+K^-$
have been predicted in Ref. \cite{Chen:2013wca}. By studying the
hidden-charm dipion decays of the charmoniumlike state $Y(4360)$ with
the ISPE mechanism, we have shown that there exist charged charmoniumlike
structures near $D\bar{D}^*$ and $D^*\bar{D}^*$ thresholds in the
$J/\psi\pi^+$, $\psi(2S)\pi^+$ and $h_c(1P)\pi^+$ invariant mass
spectra of the corresponding hidden-charm dipion decays of $Y(4360)$
\cite{Chen:2013bha}. The ISPE mechanism has been applied to the processes
$\psi(4160)/\psi(4415)\to \pi D^{(*)}\bar{D}^{(*)}$ to predict the
enhancement structures near the thresholds of $D^\ast
\bar{D}$ and $D^\ast \bar{D}^\ast$ \cite{Chen:2012yr}. Very
recently, the BESIII Collaboration has announced another charged
charmoniumlike structure $Z_c(4025)$ in the recoil mass spectrum of
$e^+e^-\to (D^*\bar{D}^*)^\pm\pi^\mp$ at $\sqrt{s}=4.26$ GeV
\cite{Ablikim:2013emm}.

These novel phenomena of charged charmoniumlike structures have a
common peculiarity, i.e., all of them are either isovectors or isodoublets.  In the
following, it is natural to ask whether there exists the
corresponding isoscalar charmoniumlike structure as a partner of
the predicted charged charmomiumlike structures. This question
inspires our interest in further studying isoscalar charmoniumlike
structures by choosing suitable decay processes. In addition, the investigations on the isoscalar charmonium-like structures will help us to reveal the nature of $Z_c(3900)$ observed by the BESIII and Belle collaborations.

Under the ISChE mechanism, the hidden-charm di-eta decays of higher
charmonia and charmoniumlike states can be a good platform to
search for isoscalar charmoniumlike structures since $\eta$, $K$ and
$\pi$ are chiral particles. In this work, we choose the processes,
\begin{eqnarray}
Y(4660)/\psi(4790) \to \eta\left[D^{(*)}\bar{D}^{(*)}\right]/\eta\ \left[D_s^{+(*)}{D}_s^{-(*)}\right]\to
\eta\eta  J/\psi,\nonumber
\end{eqnarray}
where both $D^{(*)}\bar{D}^{(*)}$ and $D_s^{+(*)}{D}_s^{-(*)}$ are the intermediate states of $Y(4660)$ and $\psi(4790) $ which decay into $\eta\eta  J/\psi$. As a vector charmoniumlike state, $Y(4660)$ was reported by Belle in the $\psi(2S)\pi^+\pi^-$ invariant mass spectrum of  $e^+e^-\to \psi(2S)\pi^+\pi^-$ \cite{Wang:2007ea}, which was later confirmed by BaBar \cite{Lees:2012pv} in the same process.
$\psi(4790)$ is a predicted charmonium with a quantum number $n^{2s+1}L_J=5^3S_1$, which is derived from  the analysis of the experimental data with the resonance spectrum expansion model
\cite{vanBeveren:2008rt}. These discussed hidden-charm di-eta decays are similar to the decays of higher charmonium or charmonium-like state into $\pi^0\pi^0 J/\psi$, where the difference lies in the isospin of the intermediate $D^{(*)}\bar{D}^{(*)}$.

{In this work, we consider only the hidden-charm di-eta decays resulted from the ISChE mechanism, which similarly provides the "signal" contribution. This study is the first step of the whole study similar to our former work of charged charmonium-like structures near the $D\bar D^*$ and $D^*\bar{D}^*$ thresholds in Ref. \cite{Chen:2011xk}. 
In reality, there exist different contributions from different final state interactions when studying the hidden-charm di-eta decay, which are equivalent to the "background" contribution. The final result derived from interference between "signal" and "background" contributions gives the real invariant mass distributions of $J/\psi\eta$, where the total amplitude is expressed as
$A_{total}=A_{"signal"}+e^{i\phi}A_{"background"}$. Here
to describe this interference the phase factor is introduced, which is crucial to understand the real invariant mass distribution. However, the phase angle reflecting the interference cannot be constrained by theory, i.e., we can construct the Lagrangians to describe the "signal" and "background" contributions  by considering Lorentz invariance and a certain symmetry, by which we further write out amplitudes $A_{"signal"}$ and $A_{"background"}$,  but the phase angle $\phi$ cannot be calculated and fixed here only from theory.
Hence, we have to wait for the corresponding experimental data. If the enough data are available in future, we can carry out the fit our prediction to the experimental data by including all contributions to the discussed hidden-charm di-eta decays, which is exactly what we have done in Ref. \cite{Chen:2013coa} (see also Ref. \cite{Chen:2011pv}). }

\begin{figure}
\centering%
\begin{tabular}{cc}
\scalebox{0.45}{\includegraphics{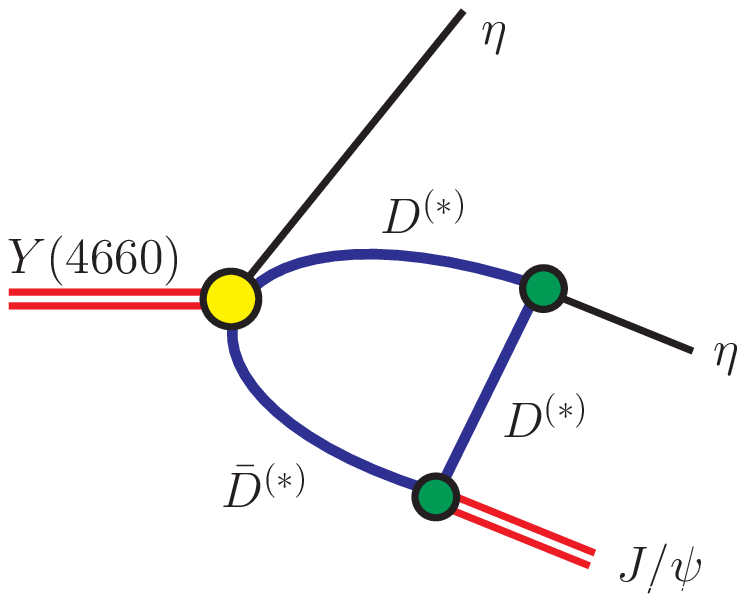}}&   %
\scalebox{0.45}{\includegraphics{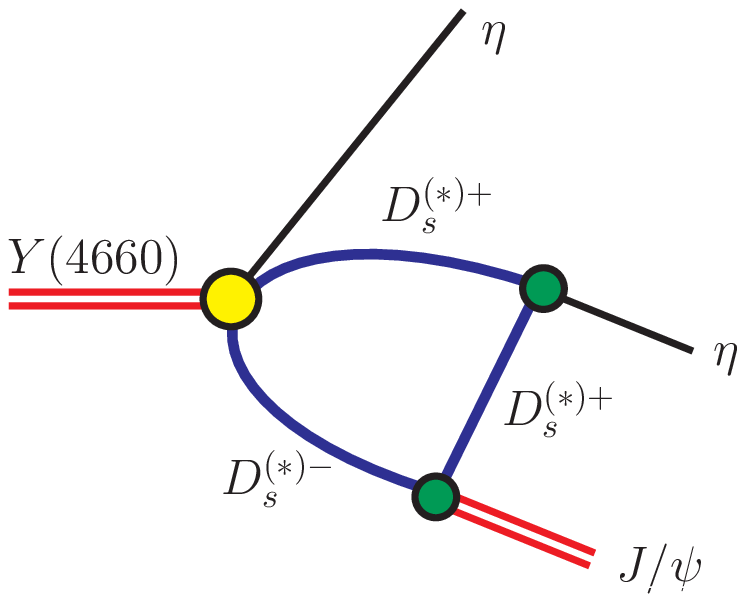}}\\  %
(a)& (b)
\end{tabular}
\caption{(Color online.) The typical diagrams for $Y(4660) \to J/\psi \eta \eta$
through the ISChE mechanism. (a) and (b) depict the processes via
charm meson and charm-strange meson loops, respectively. \label{Fig:ISChE}}
\end{figure}

Via the ISChE mechanism, the initial higher charmonium or charmoniumlike state first emits an
$\eta$ meson, which carries continuous energy distribution. Then,
the charmonium or charmoniumlike state can dissolve into the intermediate
$D^{(*)}\bar{D}^{(*)}$ and $D_s^{+(*)}{D}_s^{-(*)}$. Due to the continuous energy distribution of the emitted $\eta$, the intermediate $D^{(*)}\bar{D}^{(*)}$ and $D_s^{+(*)}{D}_s^{-(*)}$ with low momenta can easily transit into $J/\psi\eta$ by exchanging a proper charm or charm-strange meson. Taking $Y(4660) \to \eta \eta  J/\psi $ as an example, we present the corresponding typical diagrams in Fig. \ref{Fig:ISChE}.

To calculate the diagrams displayed in Fig. \ref{Fig:ISChE}, we adopt the
effective Lagrangian approach together with heavy quark limit and chiral
symmetry. The effective Lagrangian describing interactions of $Y(4660)/\psi(4790)$ with $\eta D^{(*)}\bar{D}^{(*)}$ or $\eta D_s^{+(*)} D_s^{-(*)}$ is
\cite{Kaymakcalan:1984,Haglin:2000,Oh:2000qr,Casalbuoni:1996pg,Colangelo:2002mj}
\begin{eqnarray}
\mathcal{L}_{Y \mathcal{D}^{(*)} \mathcal{D}^{(*)} \mathcal{P}}&=&
-ig_{Y \mathcal{D}\mathcal{D} \mathcal{P}} \varepsilon^{\mu \nu
\alpha \beta} Y_{\mu} \partial_{\nu} \mathcal{D}
\partial_{\alpha} \mathcal{P} \partial_{\beta} \bar{\mathcal{D}} \nonumber\\&&
+ g_{Y \mathcal{D}^\ast \mathcal{D} \mathcal{P}} {Y}^{\mu} (\mathcal{D} \mathcal{P}
\bar{\mathcal{D}}^\ast_{\mu} + \mathcal{D}^\ast_{\mu} \mathcal{P} \bar{\mathcal{D}}) \nonumber\\
&&-ig_{Y \mathcal{D}^\ast \mathcal{D}^\ast \mathcal{P}}
\varepsilon^{\mu \nu \alpha \beta} Y_{\mu} \mathcal{D}^\ast_{\nu}
\partial_{\alpha} \mathcal{P} \bar{\mathcal{D}}^\ast_\beta \nonumber\\&&-ih_{Y
\mathcal{D}^\ast \mathcal{D}^\ast \mathcal{P}} \varepsilon^{\mu \nu
\alpha \beta} \partial_{\mu} Y_{\nu} \mathcal{D}^\ast_{\alpha}
\mathcal{P} \bar{\mathcal{D}}^\ast_{\beta},
\end{eqnarray}
where $Y$ denotes a vector charmonium $\psi(4790)$ or
charmoniumlike state $Y(4660)$ and
$\mathcal{D}^{(\ast)}=\left(D^{(\ast)0}, D^{(\ast)+}, D_s^{(\ast)+ }\right)$.
$\mathcal{P}$ is the matrix representation of the pseudoscalar mesons.
Considering the $\eta$ and $\eta^\prime$ meson mixing, one has
$\mathcal{P}$ in the form,
\begin{eqnarray}
  \mathcal{P} &=&
 \left(
 \begin{array}{ccc}
\frac{\pi^0}{\sqrt{2}} + \alpha \eta + \beta \eta^\prime & \pi^{+} & K^{+}\\
\pi^{-} & -\frac{\pi^0}{\sqrt{2}}+ \alpha \eta + \beta \eta^\prime   &  K^{0}\\
 K^{-} & \bar{K}^{0} & \gamma \eta + \delta \eta^\prime
 \end{array}
 \right),\label{Eq:PseudoM}
\end{eqnarray}
where
\begin{eqnarray}
\alpha= \frac{\cos \theta -\sqrt{2} \sin \theta}{\sqrt{6}},~ \beta =
\frac{\sin \theta + \sqrt{2} \cos \theta}{\sqrt{6}},\label{11}\\
\gamma = \frac{-2 \cos \theta -\sqrt{2} \sin \theta}{\sqrt{6}} ,~
\delta= \frac{-2 \sin \theta + \sqrt{2} \cos \theta}{\sqrt{6}}\label{22}
\end{eqnarray}
and in the present work we adopt $\theta=-19.1^\circ$, which is
determined from the decay of $J/\psi$
\cite{Coffman:1988ve,Jousset:1988ni}. The concrete values of the
coupling constants $g/h_{Y\mathcal{D}^{(\ast)} \mathcal{D}^{(\ast)}
\mathcal{P}}$ are strongly dependent on the internal structures of
$Y(4660)$ and $\psi(4790)$ and we adopt $g_{Y\mathcal{D}^\ast
\mathcal{D}^\ast \mathcal{P}}=h_{Y\mathcal{D}^\ast \mathcal{D}^\ast
\mathcal{P}}$, which holds in the heavy quark limit. Since in this work we only concentrate on
the lineshape of the $J/\psi\eta$ invariant mass spectrum, the corresponding lineshape is not dependent on the overall value of the coupling
$g_{Y\mathcal{D}^{(\ast)} \mathcal{D}^{(\ast)} \mathcal{P}}$.

In addition, the interaction between $J/\psi$ and charm/charm-strange mesons can be
constructed in the heavy quark limit, which has the form
\begin{eqnarray}
\mathcal{L}_{{J/\psi \mathcal{D}^{(\ast)}\mathcal{D}^{(\ast)}}}&=& i
g_{{J/\psi \mathcal{D}\mathcal{D}}} \psi_\mu \left(
\partial^\mu \mathcal{D} {\mathcal{D}}^{\dagger} - \mathcal{D}
\partial^\mu {\mathcal{D}}^{\dagger}
\right) \nonumber\\&&-g_{{J/\psi \mathcal{D}^* \mathcal{D}}}^{}
\varepsilon^{\mu\nu\alpha\beta}
\partial_\mu \psi_\nu \left(
\partial_\alpha \mathcal{D}^*_\beta {\mathcal{D}}^{\dagger}
+ \mathcal{D} \partial_\alpha {\mathcal{D}}^{*\dagger}_\beta \right)
\nonumber\\
&& -i g_{{J/\psi \mathcal{D}^\ast \mathcal{D}^\ast}} \Big\{ \psi^\mu
\big(
\partial_\mu \mathcal{D}^{*\nu} {\mathcal{D}}_\nu^{*\dagger} -
\mathcal{D}^{*\nu}
\partial_\mu {\mathcal{D}}_\nu^{*\dagger} \big) \nonumber\\
&& + \left( \partial_\mu \psi_\nu \mathcal{D}^{*\nu} - \psi_\nu
\partial_\mu \mathcal{D}^{*\nu} \right) {\mathcal{D}}^{*\mu\dagger}  \mbox{} \nonumber\\
&& + \mathcal{D}^{*\mu}\big( \psi^\nu
\partial_\mu {\mathcal{D}}^{*\dagger}_{\nu} - \partial_\mu \psi_\nu {\mathcal{D}}^{*\nu\dagger}
\big) \Big\}.
\end{eqnarray}
In the heavy quark limit, the coupling constants satisfy the relation
\begin{eqnarray}
g_{J/\psi D D} &=& g_{J/\psi D^\ast D}\sqrt{m_{D^\ast} m_{D}} =
g_{J/\psi D^\ast D^\ast} \frac{m_{D}}{m_{D^\ast}}
=\frac{m_{J/\psi}}{f_{J/\psi}},\nonumber\\
g_{J/\psi D_s^{(\ast)} D_s^{(\ast)}} &=& \sqrt{{m_{D_s^{(\ast)}}
m_{D_s^{(\ast)}}}/{m_{D^{(\ast)}} m_{D^{(\ast)}}}} g_{J/\psi
D^{(\ast)} D^{(\ast)}},
\end{eqnarray}
where $f_{J/\psi}=416$ MeV is the decay constant of $J/\psi$, which
can be evaluated by the leptonic decay width of $J/\psi$
\cite{Beringer:1900zz}.

 Considering chiral symmetry and heavy quark
limit, we also have
\begin{eqnarray}
\mathcal{L}_{\mathcal{D}^{(\ast)}\mathcal{D}^{(\ast)} \mathcal{P}}
&=& -i g_{\mathcal{D}^*\mathcal{D} \mathcal{P}} (\bar{\mathcal{D}}
\partial_\mu \mathcal{P} \mathcal{D}^{*\mu}  -
\bar{\mathcal{D}}^{*\mu}  \partial_\mu
\mathcal{P}  \mathcal{D} ) \nonumber\\
&& + \frac{1}{2} g_{\mathcal{D}^*\mathcal{D}^*
\mathcal{P}}\epsilon_{\mu\nu\alpha\beta} \bar{\mathcal{D}}^{*\mu}
\partial^\nu \mathcal{P}\;  {\stackrel{\leftrightarrow}{\partial^\alpha}}\;
\mathcal{D}^{*\beta} ,
\end{eqnarray}
where ${\stackrel{\leftrightarrow}{\partial^\alpha}}$ only operates on
$\mathcal{D}^{*}$ and $\bar\mathcal{D}^{*}$ and the relevant coupling constants satisfy
 $g_{D^\ast D^\ast  \mathcal{P}}=
g_{D^\ast  D \mathcal{P}}/\sqrt{m_{D} m_{D^\ast}} =2 g/f_\pi $ and
$g_{D_s^{(\ast)} D_s^{(\ast)} \mathcal{P}} = \sqrt{{m_{D_s^{(\ast)}}
m_{D_s^{(\ast)}}}/{m_{D^{(\ast)}} m_{D^{(\ast)}}}} g_{D^{(\ast)}
D^{(\ast)} \mathcal{P}}$, where $f_{\pi}=132$ MeV is the pion decay
constant and $g=0.59$ is estimated from the partial decay
width of $D^\ast \to D \pi$ \cite{Beringer:1900zz}. { 
In the effective coupling, the isospin factors of the pseudoscalar mesons in the matrix should be involved, such as, $g_{D^{(\ast)} D^{(\ast)} \eta} =\alpha g_{D^{(\ast)} D^{(\ast) } \mathcal{P}}$ and $g_{D_s^{(\ast)} D_s^{(\ast)} \eta} =\gamma g_{D_s^{(\ast)} D_s^{(\ast) } \mathcal{P}}$, where $\alpha$ and $\gamma$ are defined in Eqs. (\ref{11})-(\ref{22}), which are related to the mixing angle between $\eta$ and $\eta^\prime$.  }

With the above effective Lagrangian, we can obtain the amplitudes
corresponding to the diagrams in Fig. \ref{Fig:ISChE}. In the
following, we adopt a symbol $\mathcal{M}_{AB}^{C}$ to represent the amplitude of this process, i.e.,
the initial charmonium/charmoniumlike state dissolves into
a meson pair $AB$ with one $\eta$ emission, which transits into $\eta J/\psi$ in the final state
by exchanging a meson $C$. We can express this process
$$Y(4660)(p_0) \to\eta(p_3) \left[ A(p_1)B(p_2)\right]_{C(q)}
\to\eta(p_3) \left[\eta(p_4) J/\psi(p_5) \right],$$
which is marked by the corresponding four momentum.

Taking $Y(4660) \to \eta \eta J/\psi $ via the $D\bar{D}$ intermediate
state as an example, we write out its decay amplitude, which is of the form
\begin{eqnarray}
\mathcal{A}_{D \bar{D}}^{D^\ast}  &=&(i)^3 \int \frac{d^4
q}{(2\pi)^4} \left[-ig_{Y DD \eta} \varepsilon_{\mu \rho \alpha
\beta} \epsilon_{Y}^\mu (ip_1^\rho) (ip_3^\alpha)
(ip_2^\beta)\right]
\nonumber\\
&&\times\left [ig_{D^\ast D \eta} (-ip_4^\lambda)\right]\left
[-g_{J/\psi D^\ast D} \varepsilon_{\delta \nu \theta \phi}
(ip_5^\delta)
\epsilon_{J/\psi}^\nu (-iq^\theta)\right]\nonumber\\
&&\times  \frac{1}{p_1^2-m_D^2} \frac{1}{p_2^2-m_D^2}
\frac{-g_{\lambda}{}^ {\phi} +q_{\lambda}
q^{\phi}/m_{D^\ast}^2}{q^2-m_{D^\ast}^2} \mathcal{F}^2\left(q^2,m_{D^*}^2\right),\nonumber\\\label{eq1}
\end{eqnarray}
where a form factor $\mathcal{F}\left(q^2,m_{D^*}\right)=((\Lambda^2-m_{D^*}^2)/(q^2-m_{D^*}^2))^N$ is introduced, which plays an important role to describe the off-shell effect of the exchanged charmed meson and reflect the vertex effect. In addition, the form factor also plays a role to remove divergence of the loop integral, which is similar to the Pauli-Villas renormalization scheme. We further reparameterize the cutoff $\Lambda$ as $\Lambda= \alpha_{\Lambda} \Lambda_{QCD} + m_E$, where $m_E$ is the mass of the exchanged meson and $\Lambda_{QCD}=0.22$ GeV. We adopt a typical monopole expression of a form factor, i.e., $N=1$ and take a typical parameter $\alpha_{\Lambda}=1$ to present the following numerical results.
We will later discuss the dependence of our results on the different form factors and parameter $\alpha_\Lambda$.

As indicated in Ref. \cite{Chen:2011xk}, the corresponding line shapes are not strongly dependent on $\alpha_{\Lambda}$. The total
amplitudes of $Y(4660) \to \eta \eta J/\psi $
with the $D\bar{D}$ intermediate state contribution can be expressed as
\begin{eqnarray}
\mathcal{M}_{D \bar{D}} =2 \mathcal{A}_{D \bar{D}}^{D^\ast},
\end{eqnarray}
where the factor $2$ is due to the isospin symmetry.
If considering the intermediate charm-strange meson loop contribution for
$Y(4660) \to \eta \eta J/\psi $, this factor 2 should be replaced by the factor 1.  In addition,
 the parameters in Eq. (\ref{eq1}) should be replaced with those relevant to the charm-strange meson, i.e., $m_{D^{(*)}}\to m_{D_s^{(*)}}$.

Calculating in the similar way, we can
construct the amplitudes for $Y(4660) \to \eta \eta J/\psi $ via the intermediate $(D\bar D^{\ast}+H.c.) /(D_{s}^+ D_{s}^{\ast -}+H.c.)$
and $D^\ast\bar D^{\ast} /D_{s}^{\ast+} D_{s}^{\ast-}$ (see Ref. \cite{Chen:2013wca} for more details).
Finally we obtain the general
expression of the differential decay width for $Y(4660) \to \eta
\eta J/\psi$,
\begin{eqnarray}
&&d^2\Gamma_i[Y(4660)(p_0) \to \eta(p_3) \eta(p_4) J/\psi(p_5)] \nonumber\\ &=&  \frac{|\mathcal{M}_i|^2} {3(2\pi)^3 32 m_{Y(4660)}^3} dm_{35}^2 dm^2_{45} \nonumber\\&=& \mathcal{G}_i(m_{35},m_{45}) dm_{35} dm_{45}\label{eq2}
\end{eqnarray}
with the subscripts $i=D\bar{D}, D_s\bar{D}_s, D\bar{D}^*, D_s\bar{D}_s^*, D^*\bar{D}^*, D_s^*\bar{D}_s^*$ to distinguish contributions from different intermediate states. Here, $m_{ij}^2=(p_i+p_j)^2$ and $\mathcal{G}_i=(2m_{35})(2 m_{45}) |\mathcal{M}_i|^2 /(3(2\pi)^3 32m_{Y(4660)}^3)$.
$m_{Y(4660)}$ denotes the mass of $Y(4660)$. To calculate the
$\psi(4790)\to \eta\eta J/\psi$ decay, we only need to replace the
parameters in the  above decay amplitude and differential decay
width in Eq. (\ref{eq2}).

\begin{table}[htbp]
\caption{The concrete values of the coupling constants and masses
involved in the present work. The masses are in unit of GeV.
\label{Tab:input}}
\begin{tabular}{cccccc}
\toprule[1pt] %
Coupling & Value & Coupling & Value & Coupling & Value\\
\midrule[1pt] %
$g_{J/\psi D D} $                 &  7.44               &    %
$g_{J/\psi D^\ast D} $            &  3.84 GeV$^{-1}$    &    %
$g_{J/\psi D^\ast D^\ast} $       &  8.01               \\   %
$g_{J/\psi D_s D_s} $             &  7.84               &    %
$g_{J/\psi D_s^\ast D_s} $        &  4.04 GeV$^{-1}$    &    %
$g_{J/\psi D_s^\ast D_s^\ast} $   &  8.42               \\   %
$g_{D^\ast D \eta}$               &  9.95               &    %
$g_{D^\ast D^\ast \eta}$          &  5.14 GeV$^{-1}$    &    %
$g_{D_s^\ast D_s \eta}$           & -10.62              \\   %
$g_{D_s^\ast D_s^\ast \eta}$      & -5.48 GeV$^{-1}$    &  \\  %
\bottomrule[1pt]%
\toprule[1pt] %
mass & Value & mass & Value & mass  & Value\\
\midrule[1pt] %
$m_D              $  &  1.867   &   %
$m_{D^\ast}         $  &  2.009   &   %
$m_{D_s}          $  &  1.968   \\   %
$m_{D_s^\ast}     $  &  2.112   &   %
$m_\eta           $  &  0.548   &   %
$m_{J/\psi}       $  &  3.097   \\   %
$m_{Y(4660)}      $  &  4.660   &   %
$m_{\psi(4790)}   $  &  4.790   &\\ %
\bottomrule[1pt]
\end{tabular}
\end{table}

The input parameters including the masses and coupling constants adopted in this work are listed in Table \ref{Tab:input}.

\begin{figure}[htbp]
\centering%
\scalebox{0.70}{\includegraphics{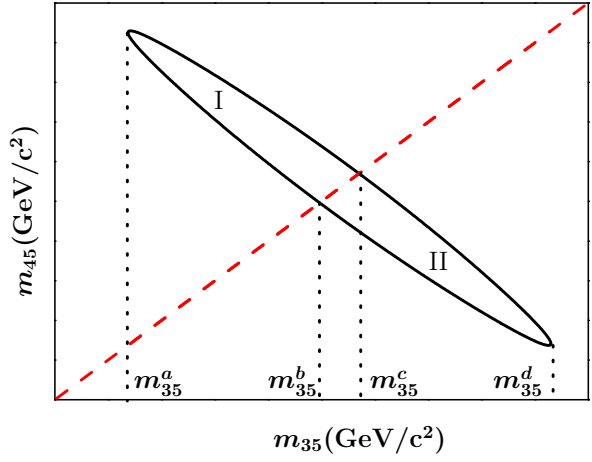}} %
\caption{(Color online.) A sketch diagram of the phase space of $Y(4660)/\psi(4790)(p_0) \to \eta(p_3) \eta(p_4) J/\psi(p_5)$ in terms of $m_{35}$ and $m_{45}$. Inside of the black solid curve is the phase space of the decay process and the diagonal dashed one represents the line of $m_{35}=m_{45}$. The minimum and maximum of $m_{35}$ are $m_{35}^a=m_3+m_5$ and $m_{35}^d=m_0-m_4$, respectively. The meanings of $m_{35}^{b}$ and $m_{35}^c$ are defined in the main text. \label{Fig:phsp} }
\end{figure}

Since two $\eta$ mesons in the final state are identical, we give the distribution in terms of $M_{\mathrm{max}}({J/\psi \eta})$ defined below, which denotes the maximum distribution of the $J/\psi\eta$ invariant mass spectrum of  $Y(4660)(p_0) \to \eta(p_3) \eta(p_4) J/\psi(p_5)$.
In Fig. \ref{Fig:phsp}, we present a sketch diagram of the phase space depending on $m_{35}$ and $m_{45}$. The phase space is divided into two parts by the diagonal line of $m_{45}=m_{35}$.

%
Considering that the distribution is symmetric in $m_{35}$ and $m_{45}$, i.e., $d^2\Gamma \left(m_{35}, m_{45} \right)=d^2\Gamma \left(m_{45}, m_{35} \right)$. In addition, as shown in Fig. \ref{Fig:phsp}, the phase space is also symmetric to the line of $m_{45}=m_{35}$.  Then the one-dimensional distribution of $M_{\mathrm{max}}({J/\psi \eta})$ can be evaluated as follows,
\begin{eqnarray}
 &&\frac{d\Gamma\left(M_\mathrm{max}\left( J/\psi\eta
\right)\right)}{dM_{\mathrm{max}}\left( J/\psi\eta \right)} \nonumber \\
 &&=2\int^{\mathrm{II}} {\cal G}\left(M_{\mathrm{max}},
m_{45}\right)dm_{45}, \label{eq:Mmax}
\end{eqnarray}
because $\int^{\mathrm{I}} {\cal G}\left(m_{35}, M_{\mathrm{max}}
\right)dm_{35}= \int^{\mathrm{II}} {\cal G}\left(M_{\mathrm{max}},
m_{45}\right)dm_{45}$, where $M_{\mathrm{max}}\left( J/\psi\eta
\right)=\mathrm{max}\left\{m_{35}, m_{45}\right\}$ of the integrand,
${\cal G}\left(m_{35},m_{45}\right)$, in each integral of r.h.s. of Eq.
(\ref{eq:Mmax}).
The minimum of $M_{\mathrm{max}}({J/\psi \eta})$ is $m_{35}^b$, which is $3.838$ GeV for $Y(4660) \to \eta \eta J/\psi$ and $3.890$ GeV for $\psi(4790) \to \eta \eta J/\psi$. One has to notice that the phase space has a sudden change at the point $m_{35}^{c}$, which will lead to a turning point in the distributions of $M_{\mathrm{max}}({J/\psi \eta})$. This turning point appears at $M_{\mathrm{max}}({J/\psi \eta})=3.918$ GeV and $M_{\mathrm{max}}({J/\psi \eta})=3.996$ GeV for $Y(4660) \to \eta \eta J/\psi$ and $\psi(4790) \to \eta \eta J/\psi$, respectively.

\begin{figure}[htb]
\centering%
\scalebox{0.53}{\includegraphics{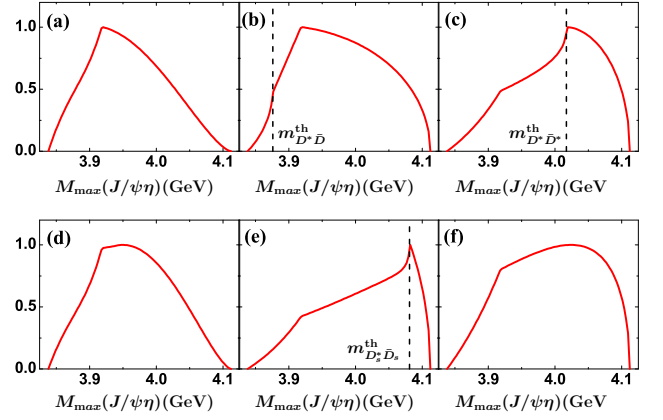}} %
\caption{(Color online.) The obtained $M_{\mathrm{max}}({J/\psi \eta})$ distributions of $Y(4660) \to \eta \eta J/\psi$.
Here, the diagrams (a), (b) and (c) are
the lineshapes resulted from the intermediate $D\bar{D}$, $D^\ast
\bar{D} +H.c.$ and $D^\ast \bar{D}^\ast$, respectively, while the dagrams (d), (e) and (f) are the results considering the intermediate $D_s\bar{D}_s$, $D_s^\ast \bar{D}_s +H.c.$ and $D_s^\ast
\bar{D}_s^\ast$ contributions, respectively. The thresholds of $D\bar{D}^*$, $D^*\bar{D}^*$ and $D_s^*\bar{D}_s$ are marked by the vertical dashed lines. The maxima of these lineshapes are normalized to 1.
\label{Fig:Y4660} }
\end{figure}

\begin{figure}[htb]
\centering%
\scalebox{0.53}{\includegraphics{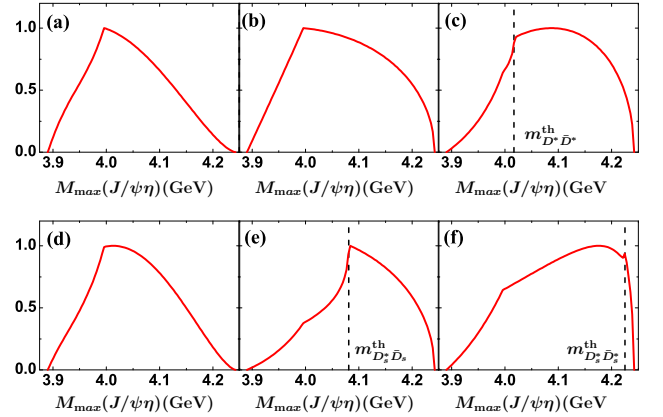}} %
\caption{(Color online.) The obtained $M_{\mathrm{max}}({J/\psi \eta})$
distributions of $\psi(4790) \to \eta \eta J/\psi$.  The lineshapes
presented here are arranged in the same way as in Fig.
\ref{Fig:Y4660}.}\label{Fig:Y4790}
\end{figure}

Separately considering the different intermediate
state contributions, the calculated results of $Y(4660) \to \eta \eta J/\psi $ and $\psi(4790)\to
\eta\eta J/\psi$ are shown in Figures
\ref{Fig:Y4660} and \ref{Fig:Y4790}, respectively. As presented in Figures \ref{Fig:Y4660} and \ref{Fig:Y4790}, the
obtained lineshapes are not smooth since there exist the turning points in
all the diagrams, which appear when
$M_{{\mathrm{max}}}(J/\psi\eta)=$ 3.918 GeV and
$M_{{\mathrm{max}}}(J/\psi\eta)=$ 3.996 GeV for the $Y(4660) \to \eta \eta
J/\psi $ and $\psi(4790) \to \eta \eta J/\psi$ decays, respectively. These
turning points are due to the maximum distribution of the
$J/\psi\eta$ invariant mass spectrum itself rather than the ISChE
mechanism.

As for $Y(4660) \to \eta \eta J/\psi$, we present the following important information:
\begin{enumerate}
\item Besides the turning points mentioned above, there do not exist other peaks in Fig. \ref{Fig:Y4660} (a) and (d), which is the same as the results for dipion decay cases. No peak caused by ISChE mechanism appears in Fig. \ref{Fig:Y4660} (f), since the threshold of $D_s^\ast \bar{D}_s^\ast$ is 4224.6 MeV, that is beyond the phase space.
\item FIg. \ref{Fig:Y4660} (b) indicates the existence of another peak appearing in the $M_{\mathrm{max}}({J/\psi \eta})$ distribution caused by ISChE mechanism with $D^\ast \bar{D}$ intermediate states.
\item There are explicit enhancement structures in Figs. \ref{Fig:Y4660} (c) and (e). As a broad structure, the enhancement in Fig. \ref{Fig:Y4660} (c) is around $M_{\mathrm{max}}({J/\psi \eta})=m^{\rm th}_{D^*\bar D^*}=4.018$ GeV. While, the enhancement in Fig. \ref{Fig:Y4660} (e) is a sharp peak at $m^{\rm th}_{D^*_s\bar D_s}=4.081$ GeV.
\end{enumerate}

Similar to the above analysis of $Y(4660) \to \eta \eta J/\psi$, in
the following we also have some extra novel phenomena of $\psi(4790) \to
\eta \eta J/\psi$, which include:
\begin{enumerate}
\item There is an enhancement structure near $M_{\mathrm{max}}({J/\psi
\eta})=4.081$ GeV as shown in Fig. \ref{Fig:Y4790} (e). In
addition, a small peak appears in the $M_{\mathrm{max}}({J/\psi
\eta})$ distribution in Fig. \ref{Fig:Y4790} (f). We also notice a peak due to a threshold
at $M_{\mathrm{max}}({J/\psi \eta})=4.018$ GeV in Fig.
\ref{Fig:Y4790} (c), which is resulted from the ISChE mechanism and
is different from the turning point at
$M_{\mathrm{max}}({J/\psi \eta})=3.996$ GeV mentioned above.
\item The lineshapes listed in Figs. \ref{Fig:Y4790} (a), (b) and (d)
show that the intermediate $D\bar{D}$, $D\bar{D}^*$, $D_s\bar{D}_s$
cannot result in enhancement structures in the corresponding
$M_{\mathrm{max}}({J/\psi \eta})$ distributions except for peaks at the turning points.
\end{enumerate}

\begin{figure}[htb]
\centering%
\scalebox{0.35}{\includegraphics{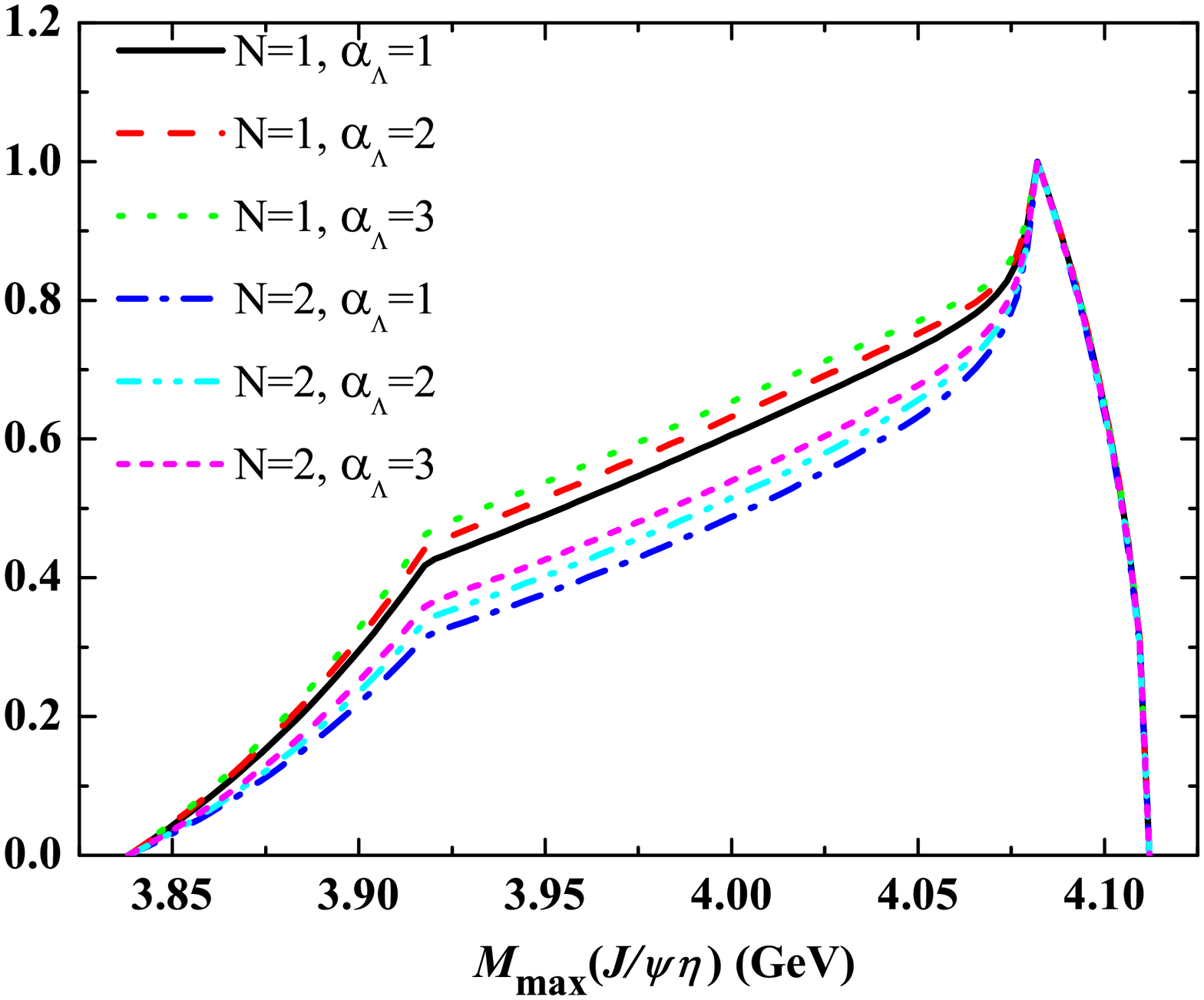}} %
\caption{(Color online.) The lineshapes of $d\Gamma(\psi(4660) \to \eta \eta J/\psi)/ dM_{max}(J/\psi \eta)$ 
dependent on dipole and monopole form factors and parameter $\alpha_\Lambda$. Here, the figure is resulted from the ISChE mechanism considering only the $D_s^\ast \bar{D}_s +H.c.$ intermediate state. The maxima of the lineshapes are normalized to be unity.}\label{Fig:ffdep}
\end{figure}

{In the following, we discuss the form factor dependence of the lineshapes obtained, where we consider monopole and dipole form factors in the calculation. In Fig. \ref{Fig:ffdep},
we present the distribution of $d\Gamma(\psi(4660) \to \eta \eta J/\psi)/ dM_{max}(J/\psi \eta)$ obtained by taking these two form factors, and compare their results, which show that the peak structures are not strongly dependent on the expressions of the form factor. In addition, we also study  $\alpha_\Lambda$ dependence of the distribution of $d\Gamma(\psi(4660) \to \eta \eta J/\psi)/ dM_{max}(J/\psi \eta)$. It is shown in Fig. \ref{Fig:ffdep} that the lineshapes for $d\Gamma(\psi(4660) \to \eta \eta J/\psi)/dM_{max}(J/\psi \eta)$ are not sensitive to $\alpha_\Lambda$, which is consistent with the observation in Ref. \cite{Chen:2011xk}, where we once calculated $\alpha_\Lambda$ dependence of the lineshapes of $d{\Gamma(\psi(4415 \to \pi^+ \pi^- h_c))}/dm_{h_c \pi^+}$ for the process $\psi(4415) \to \pi^+ \pi^- h_c$ as an example, which also indicates that the corresponding lineshapes are weakly dependent on the cutoff introduced in the form factor. Accordingly we can conclude that the form factors and the parameter $\alpha_{\Lambda}$ weakly affect the obtained lineshapes, especially in the vicinity of $D_s^\ast \bar{D}_s$ threshold. }

In summary,  we have calculated the di-eta decay of the higher charmonia via the ISChE mechanism and predicted some enhancements around the thresholds of  $\mathcal{D} \bar{\mathcal{D}}^{\ast}$ and $\mathcal{D}^{\ast} \bar{\mathcal{D}}^{\ast}$. Owing to the $J^{PC}$ conservation, $\mathcal{D} \bar{\mathcal{D}}$ should be in $P-$wave, while $\mathcal{D} \bar{\mathcal{D}}^{\ast}$ and $\mathcal{D}^{\ast} \bar{\mathcal{D}}^{\ast}$ are in $S-$wave in the $\psi \mathcal{D}^{(\ast)} \mathcal{D}^{(\ast)} \eta$ effective couplings. The stronger $S$-wave couplings mainly contribute to the ISChE mechanism, which may be the reason why there is no enhancement around $\mathcal{D} \bar{\mathcal{D}}$ threshold via the ISChE mechanism.

In the past decade experiments have made big progress on
searching for charmoniumlike states $XYZ$, which also stimulated
extensive theoretical studies on their properties. At present,
it is still a hot and interesting topic to carry out both theoretical
and experimental investigations on these $XYZ$ states.

Recent experimental observation of $Z_c(3900)$ is a charged
charmoniumlike state reported by BESIII \cite{Ablikim:2013mio} and
confirmed by Belle \cite{Liu:2013dau} and Ref. \cite{Xiao:2013iha}.
It again draws our attention to a charmoniumlike state since a charged
enhancement structure near the $D\bar{D}^*$ threshold was predicted
in Ref. \cite{Chen:2011xk} before this experimental observation,
where the special mechanism (ISPE) was applied to study the hidden-charm
dipion decay of $Y(4260)$ and other higher charmonia. Our prediction
confirmed by BESIII also inspires our interest in applying and
extending the ISPE to provide more abundant phenomena of charged
charmoniumlike structures
\cite{Chen:2013bha,Chen:2012yr,Chen:2013wca}.

Although we already have given many predictions of charged charmoniumlike
structures \cite{Chen:2013bha,Chen:2012yr,Chen:2013wca}, we notice
that the isoscalar charmoniumlike structures similar to the predicted
charged ones are absent in experiment. Thus,
in this work we have studied the hidden-charm di-eta decays of $Y(4660)$
and $\psi(4790)$ to theoretically give the prediction of the
isoscalar charmoniumlike structure. Our results show that there are
enhancement structures near $D\bar{D}^*$, $D^*\bar{D}^*$ and
$D_s\bar{D}_s^*$ thresholds in the $M_{\mathrm{max}}({J/\psi \eta})$
distribution of $Y(4660) \to \eta \eta J/\psi$. The calculation of
$\psi(4790) \to \eta \eta J/\psi$ predict the enhancement structure
near $D^*\bar{D}^*$, $D_s\bar{D}_s^*$ and $D_s^*\bar{D}_s^*$
thresholds in the corresponding $M_{\mathrm{max}}({J/\psi \eta})$
distributions.
We have not seen any enhancement in the intermediate $D\bar D/D_s\bar D_s$ channels in both
$Y(4660)/\psi(4790)\to \eta \eta J/\psi$ processes.
Other than thresholds, we have found the turning
points owing to the sudden change of the phase space.Because we have separately given only
lineshapes of different intermediates, we cannnot definitely claim that
experiments should find these peaks due to the interferences between different mechanisms working in the higher charmonia and charmonium-like state decays.  

These theoretical studies provide abundant
information on isoscalar charmoniumlike structure, which will be
helpful for further experimental exploration in future, where the
potential experiments to search for the predicted enhancements in
this work include BESIII, Belle, BaBar, and forthcoming BelleII.

\section*{Acknowledgement}

This project is supported by the National Natural Science
Foundation of China under Grant No. 11222547, No. 11175073, No.
11005129, No. 11375240 and No. 11035006, the Ministry of Education of China
(FANEDD under Grant No. 200924, SRFDP under Grant No.
2012021111000, and NCET), the Fok Ying Tung Education Foundation
(No. 131006), and the West Doctoral Project of Chinese Academy of
Sciences.

\end{document}